\documentclass[review]{elsarticle}

\usepackage{lineno,hyperref}
\hypersetup{colorlinks = true,linkcolor = blue,anchorcolor =red,citecolor = blue,filecolor = red,urlcolor = red,
            pdfauthor=author}

\usepackage{amsmath}
\usepackage{amsfonts,amsthm,dsfont}
\usepackage{graphicx}
\usepackage{adjustbox}
\usepackage{enumerate}
\usepackage{natbib}
\usepackage{pgf,subfig}
\usepackage{soul}
\usepackage{multirow}
\usepackage{url}
\usepackage{geometry}
\usepackage{pdflscape}
\usepackage{rotating}
\usepackage{mathtools}

\usepackage[ruled]{algorithm2e}
\SetKwInput{KwData}{Input}
\SetKwInput{KwResult}{Output}

\definecolor{green}{rgb}{0.0,0.6,0.12}

\definecolor{blue}{rgb}{0.0,0.12,0.6}

\newcommand{\R}{\mathbb{R}} 

\journal{Journal of \LaTeX\ Templates}

\begin{document}

\begin{frontmatter}

\title{Self-Organizing Maps for Exploration of Partially Observed Data and Imputation of Missing Values}


\author[mymainaddress,mysecondaryaddress]{Sara Rejeb\corref{mycorrespondingauthor}}
\cortext[mycorrespondingauthor]{Corresponding author}
\ead{sara.rejeb@sorbonne-universite.fr}

\author[mysecondaryaddress]{Catherine Duveau}

\author[mymainaddress]{Tabea Rebafka}

\address[mymainaddress]{LPSM, Sorbonne Université, Université de Paris \& CNRS, 4,
Place Jussieu, 75252, Paris cedex 05, France}
\address[mysecondaryaddress]{Safran Aircraft Engines, Réau, 77550 Moissy-Cramayel,
France}

\begin{abstract}
The self-organizing map is an unsupervised neural network which is widely used for data
visualisation and clustering in the field of chemometrics. The classical Kohonen algorithm that
computes self-organizing maps is suitable only for complete data without any missing values. However, in many applications, partially observed data are the norm. 
In this paper, we propose an extension of self-organizing maps to incomplete data via a new criterion that also defines estimators of the missing values.
In addition, an adaptation of the Kohonen algorithm, named missSOM,  is provided
to compute these self-organizing maps and impute missing values. An efficient implementation is provided. 
Numerical experiments on simulated data and a chemical dataset illustrate the short computing time of missSOM and assess its performance  regarding various criteria and in comparison to the state of the art.
\end{abstract}

\begin{keyword}
Self-organizing maps \sep partially observed data \sep missing data imputation \sep robustness to missingness mechanism 
\end{keyword}

\end{frontmatter}


\section{Introduction}

In chemometrics, data exploration is essential and is generally the first part of the analysis of any dataset. 
With the explosion of data volume in many fields of application due to big data, the structure of data is often hidden behind the massive data and difficult to detect. This increases the importance of data exploration methods providing meaningful information on the internal structure and correctness of the data, as well as the relationships and redundancies among the variables.
The initial understanding of the data gained by exploratory data analysis is particularly useful for data modeling.

Common tasks of data exploration are visualization and clustering of the data. While there is plethora of methods addressing one of the tasks, self-organizing maps  simultaneously provide both a low-dimensional visual data representation in form of a map and a clustering of the observations.
 Introduced by~\cite{kohonen}, this approach consists in mapping the data and performing  vector quantization of the input space while preserving  topological properties of the data even when these data are high-dimensional. These are   properties that other data analysis methods  do not have. For instance, $k$-means  only categorises data, but the  topology gets lost and the method does not allow data visualisation. 
 Several examples of application in different areas  are provided in~\citep{cottrell:hal-01796059}   illustrating the practical interest of self-organizing maps. 
In \cite{QIAN2019100020} it is shown that self-organizing maps are a powerful tool for visualisation of high-dimensional data. A fraud detection method based on the SOM visualization and classification is proposed in \cite{OLSZEWSKI2014324}. In~\citep{PENN2005531}, self-organizing maps are used to visualize and classify complex geologic data. Moreover, self-organizing maps have proven to be of considerable   value in  finance, where they are able to structure, analyze, and visualize large amounts of multidimensional financial data in a significant manner~\citep{deboeck2013visual, doi:10.1057/palgrave.ivs.9500048}.
In short, as self-organizing maps provide easily interpretable results with a global view of the data, they
 have become very popular 
in many fields of application. Furthermore, self-organizing maps are widely used in chemometrics~\citep{chimie4, chimie2, chimie3, chimie1, chimie5, agriculture}, but also in biology~\citep{MELIN2020109917}, humanities~\citep{massoni}, industry, such as health monitoring of aircraft engines~\citep{jerome}. 
Many variants of the standard self-organizing map have been developed, such as Generative Topographic Mapping~\citep{gtm}, which is a probabilistic version of the self-organizing map, or extensions to more complex data types (mixed, textual, etc.)~\citep{mixte, Lebbah2005MixedTM} demonstrating the relevance of self-organizing maps until today.

A common issue with datasets in most fields of application are missing data. Data may be incomplete  for a large variety of reasons. In surveys, for instance, they occur due to non-responses to questions that affect privacy ~\citep{survey1, survey2, MIRZAEI2021}.
In industrial applications and chemometrics, measuring instruments may have malfunctions or detection limits yielding erroneous and missing entries~\citep{industry1, industry2}. In medical research, missing data can occur in clinical trials when patients abandon or stop taking the treatment for a certain period of time~\citep{medical, medical2, medical3}. Missing data   also frequently occur in chemical~\citep{missingChimie, missingChimie2} and environmental~\citep{missingEnvironnement} studies.
Moreover, concerning huge databases, merging several datasets from different sources can also result in missing data, as some entries may not be recorded at all for some of the sources. 

The impact of missing data on statistical results can be serious, leading to biased estimates, loss of information, decreased statistical power, increased standard errors, and weakened generalizability of findings. 
However, for a long time, in statistics, missing data have been  treated in a very simple and inappropriate way, either by deletion of incomplete measurements or by basic data completion by mean or median values. 
This has changed during the last decades, by the development of 
 many statistical methods that  account for missing data in a meaningful way.
 There are two general approaches to deal with missing data: either a statistical method is directly adapted to the partially observed data, or first an appropriate imputation method is applied to complete the data such that the statistical method of interest can be used on  the completed or augmented data. 
 
In this paper we are interested in self-organizing maps in the presence of missing data. This problem has been considered among others by~\cite{Cottrell2007MissingV} by simply restricting all vector calculations to the observed entries. As such, all  observed data entries are taken into account in  the algorithm. However, the method  performs rather poorly  when the number of incomplete observations with multiple missing entries is large. 
Moreover, the imputation of missing data is done afterwards by replacing missing entries by the closest features on the learned map.
Related approaches are presented in~\citep{FOLGUERA2015146, SOMissing, SOMissing2, SOMissing3, SOMissing4}  and more recently in ~\citep{junno2019predicting, SOMmissing6}.
 
Missing data imputation can be useful to avoid incomplete data, which is crucial in data mining when methods cannot handle any missing entries.
We have the ambition to combine the tasks of imputation and learning the map by a principled approach. Our motivation is the fact that  any non trivial imputation method is based on some data model, and so it is natural to use the self-organizing map for imputation. Conversely, a better map may be learned when data are complete. Thus, treating both tasks simultaneously may be beneficial for the two of them. 

Our approach can be viewed as an extension of the standard Kohonen  algorithm for self-organizing maps and the principle of our method is given in  Section~\ref{sec:missSOM_short}. 
A mathematical presentation of the method,  a new loss function that encodes our double goal of imputation and learning a self-organizing map  and two algorithmic solutions are given in Section \ref{sec:missSOM}. 
Moreover, Section \ref{sec:numericalStudy} provides an extensive numerical study assessing the robust performance of our method in various settings and in comparison to alternative methods from the literature.  

\section{The new method in a nutshell} \label{sec:missSOM_short}
This section presents the principle of our new method for self-organizing maps for partially observed data. 
To start with, we recall the classical self-organizing map. 
Let $x_1,\dots,x_n$ 
be $n$ observations or measurements   of dimension $p$. 
A self-organizing map represents a nonlinear  projection of the high-dimensional data onto a low-dimensional subspace. 
This subspace is   a two-dimensional map represented as a regular grid composed of $K$ fixed neurons.  
This fixed spatial arrangement of the neurons on the map is the key for the  preservation of the topology of the input data when projected onto the map.
Every neuron $k$ is associated with a $p$-dimensional prototype vector $w_k$, also called code vector, that  is to be learned. 
The prototype vectors define a discretization of the data space, and each observation $x_i$ is assigned to its closest prototype.
Ideally, prototype vectors of neighboring neurons on the grid are close one to another, so that data points $x_i$ that are close in the input space are also close on the map.

The 
 Kohonen  algorithm computes the self-organizing map for complete data in an iterative fashion. 
One randomly picked observation $x_i$ is treated at each iteration. 
First, the winning neuron or best matching unit is determined, which is the neuron whose prototype is the closest to measurement $x_i$.
Then, all code vectors $w_k$ are updated by attracting them towards the measurement $x_i$. 
 The attraction is the strongest  for the winning neuron and  very weak for the neurons that are far from the winning neuron. 
 Those updates eventually result in an ordered map, where neighboring neurons have similar prototype vectors. 
 
Now, when some of the measurement vectors $x_i$ contain missing entries, the Kohonen algorithm is not applicable anymore. We propose to learn the missing entries while learning the map in the following way. Our algorithm is the following: starting from some initial imputed values, like the mean values computed over the observed entries, select every measurement $x_i$ once, determine its winning neuron by considering the distance only over the observed entries of $x_i$ and then update the code vectors just as in 
the classical Kohonen algorithm. Then, perform an update of the imputed values using a weighted means of the closest code vectors of the partially observed measurement. Repeat this procedure until convergence. A full description is provided in Algorithm~\ref{algo:accelmissSOM}.

This algorithm
performs both data visualization and imputation of the incomplete data.
Interestingly, it is as fast as the standard Kohonen algorithm, since the update of the imputed values is immediate. 

\section{Self-organizing maps with incomplete data} \label{sec:missSOM}

  \begin{algorithm}[t]
   \KwData{Data matrix $X$, size and topology of the map, neighborhood function $V_\lambda$, sequence of radii $(\lambda_t)_{0 \leq t \leq T}$ and learning steps $(\varepsilon_t)_{0 \leq t \leq T}$.}
   Initialize code vectors $W^{(0)}$ \;
   Initialize the counter of iterations: $t=0$ \;
  \While{not converged}{
  Increment $t$: Set $t=t+1$ \;
    Choose an observation $i\in\{1,\dots,n\}$ randomly \;
    Assignment: Compute winning neuron $\ell=h(x_i,  W^{(t-1)})$ \;
    Update code vectors: \\
   \For{$k=1,\dots,K$}{$w^{(t)}_k = w^{(t-1)}_k + \varepsilon_tV_{\lambda_t}(k, \ell)\left(x_i-w^{(t-1)}_k\right).$}
  }
  \KwResult{Code vectors $W^{(t)}$. }
     \caption{Standard Kohonen algorithm }\label{algo:standardSOM}
\end{algorithm}

In this section we  first formally state  the classical self-organizing map, before introducing the new loss function  and two algorithms for the computation of self-organizing maps with partially observed data and missing data imputation.

\subsection{Classical Kohonen algorithm} 
The data matrix containing the measurements is denoted by $X=[x_1,\dots,x_n] \in \mathbb R^{n \times p}$. 
The arrangement of the neurons on the map  is  given by some neighborhood function
$V_\lambda : \{1, \dots, K\}^2 \mapsto 
 \mathbb R_+$.
The neighborhood radius $\lambda>0$ describes the zone of influence around a neuron.
The best prototype vectors of the self-organizing map are defined as the minimum of the loss function $F$ defined by
\begin{equation}\label{eq:lossSOM}
F(W) = \frac{1}{2n}\sum_{i=1}^{n}\sum_{k=1}^{K}V_\lambda(k, h(x_i,W))\|x_i - w_k\|^2_2,
\end{equation}
where $W=[w_1,\dots,w_K]\in\R^{p\times K}$ is the matrix of $K$ prototype vectors and 
$h : \mathbb R^p \times \R^{p\times K} \mapsto
 \{1, \dots, K\}$ denotes the allocation function, attributing the closest prototype to a data point $x$ w.r.t. the Euclidean distance, defined as 
\begin{equation}\label{eq:winner}
h(x,W)=\arg\min_{1\le k\le K}\|x-w_k\|_2.
\end{equation}
The loss $F$ takes into account all distances between every measurement and all code vectors, weighted by the neighborhood function evaluated on the corresponding neurons. As a result, code vectors that minimize the loss are similar if they are close on the map.
In the specific case where the neighborhood function satisfies $V_\lambda(k,\ell)=0$ for all $k\neq\ell$, the loss $F$ is the criterion minimized by the $k$-means algorithm. That is,   clusters obtained by $k$-means are independent, while prototypes of a  self-organizing map  are organized in a topological way.

To compute the minimum of loss $F$, 
\cite{ritter} showed that in the given framework 
a gradient descent algorithm can be used, referred to as
the Kohonen stochastic algorithm. 
For a randomly picked observation $x_i$ with winning neuron $\ell=h(x_i, W^{(t)})$,
the updates of the code vectors are given by
\begin{equation}
\label{eq:updatePrototype}
w^{(t+1)}_k = w^{(t)}_k + \varepsilon_tV_{\lambda}(k, \ell)(x_i-w^{(t)}_k),
\end{equation}
 where $(\varepsilon_t)_{t\geq0}$ is a sequence of decreasing learning steps. This update  attracts all prototypes towards   observation $x_i$. 
 It is also common to shrink the neighborhood by using a decreasing  sequence of radii $(\lambda_t)_{t\geq 0}$ in the neighborhood function $V_\lambda$. The algorithm is summarized in Algorithm \ref{algo:standardSOM}.

\subsection{Notation for incomplete data} 
Now we   consider an  incomplete $n\times p$ data matrix containing missing values. 
 Let the  matrix $M=(m_{i,j})_{i,j} \in \{0, 1\}^{n \times p}$ be the missing-data pattern which indicates where the  entries     are missing  or masked, and that is defined by
\begin{equation*}
    m_{i,j}=  \left\{
\begin{array}{ll}
1 & \mbox{if } x_{i, j} \mbox{ is observed} \\
0 & \mbox{if } x_{i, j} \mbox{ is  missing} 
\end{array}
\right.
\end{equation*}

We denote $X^{\mathrm{obs}}$ the set of observed data values and  $X^{\mathrm{miss}}$ the set of non-observed data entries hidden by the missing-data pattern $M$.  The complete data are denoted by  $X^{\mathrm{compl}}=(X^{\mathrm{obs}}, X^{\mathrm{miss}})$.  Likewise,  for the observation vector  $x_i$ we denote by $x^{\mathrm{obs}}_i$ and $x^{\mathrm{miss}}_i$  the observed and unobserved entries, respectively, and, with some abuse of notation, $x_i^{\mathrm{compl}}=(x^{\mathrm{obs}}_i, x^{\mathrm{miss}}_i)$ is the complete vector, which also corresponds to the $i$-th row of $X^{\mathrm{compl}}$.

Our goal is to adapt the model of self-organizing maps to partially observed data, and moreover, learn the values of the missing data. 
The motivation to treat these tasks simultaneously is that learning missing values requires a data model, and as we are interested in self-organizing maps it is natural to use this model for data imputation. At the same time, learning a map with completed data may give better results  compared to using only  the observed part $X^{\mathrm{obs}}$ of the data. 
 
\subsection{New loss function}\label{subsec:newloss}

We introduce a new loss function that considers both problems : finding the best self-organizing map and the best values for imputation of the missing data. In other words, by minimizing the new loss function $F_{\mathrm{missom}}$
 we  search for both the best code vectors $W \in \mathbb R^{p\times K}$ for the map and the best values   for  the missing    data denoted by  
$X^{*}$ chosen in the set of all possible values for the missing entries $\mathcal X^{\mathrm{miss}}$. 

To define the new criterion, an adaptation of the definition of the winning neuron is in order.
 In the presence of missing values, 
it is  natural   to restrict the Euclidean distance  in~\eqref{eq:winner} only to the observed entries. More precisely, for any vectors $x^{\mathrm{obs}}\in\R^{p'}$ ($p'\leq p$), $m\in\{0,1\}^p$ with $\sum_{j=1}^pm_j=p'$ and code vectors $W\in \R^{p\times K}$, we set
$$
h^{\mathrm{miss}}(x^{\mathrm{obs}}, m, W)
=\arg\min_{1\le k\le K} \|x^{\mathrm{obs}} -w_{k}\odot m\|_2,
$$
where $w_{k}\odot m$ denotes the $p'$-vector made of the elements $w_{k,j}$  of $w_k$ such that $m_j=1$. 
Now, we define the new loss as

\begin{equation*}
F_{\mathrm{missom}}(W, X^{*}) = \frac{1}{2n}\sum_{i=1}^{n}\sum_{k=1}^{K} V_\lambda\left(k, h^{\mathrm{miss}}(x_i^{\mathrm{obs}}, m_i,W) \right) \left\| (x_i^{\mathrm{obs}},x_i^*) -  w_k\right\|^2_2,
\end{equation*}

where  $m_i \in \mathbb R^p$ is the $i$-th row of the matrix $M$ and $(x_i^{\mathrm{obs}},x_i^*)$ denotes the $i$-th measurement vector completed with  $x_i^*$.
Since
\begin{equation*}
    \left\| (x_i^{\mathrm{obs}},x_i^*) -  w_k\right\|^2_2 = 
    \left\| x_i^{\mathrm{obs}} -  w_k\odot m_i\right\|^2_2+ 
\left\| x_i^*   -  w_k\odot(\mathbf 1_p -m_i) \right\|^2_2,
\end{equation*}

where $\mathbf{1}_p=(1, \dots, 1)^T \in \mathbb R^p$,
the criterion $F_{\mathrm{missom}}$ can be decomposed into two parts according to the observed and the missing entries as
\begin{equation*}
F_{\mathrm{missom}}(W,  X^{*})= F_{\mathrm{obs}}(W) + F_{\mathrm{miss}}(W,  X^{*}),
\end{equation*}
where $F_{\mathrm{obs}}(W)$ is the part of the loss over the observed entries   given by 
\begin{equation*}
F_{\mathrm{obs}}(W)
= 
\frac{1}{2n}\sum_{i=1}^{n}\sum_{k=1}^{K} V_\lambda\left(k, h^{\mathrm{miss}}(x_i^{\mathrm{obs}},m_i,W) \right) 
\left\|x_i^{\mathrm{obs}}  - w_k \odot m_i\right\|^2_2,
\end{equation*} 
and
$F_{\mathrm{miss}}(W,  X^{*})$ is the contribution of the imputed values $ X^{*}$ to the loss, defined as
\begin{equation*}
F_{\mathrm{miss}}(W,  X^{*}) = \frac{1}{2n}\sum_{i=1}^{n}\sum_{k=1}^{K} V_\lambda\left(k, h^{\mathrm{miss}}(x_i^{\mathrm{obs}},m_i,W)\right)\left\|x_i^*  - w_k \odot (\mathbf{1}_p - m_i) \right\|^2.
\end{equation*}

Note that in the complete-data case, where the missing-data pattern is $M = \mathbf{1}_{n \times p}$,  $F_{\mathrm{miss}}(W,  X^{*})=0$ for any    $W$  and any $ X^{*}$, so that the criterion $F_{\mathrm{missom}}$ is equal to the one of the classical self-organizing map, that is,  $F_{\mathrm{missom}}(W,  X^{*})=F(W)$. 

 \begin{algorithm}[t]
   \KwData{Incomplete data $X^{\mathrm{obs}}$, missing-data pattern $M$,  size and topology of the map,  neighborhood function $V_\lambda$, sequence of radii $(\lambda_t)_{0 \leq t \leq T}$ and learning steps $(\varepsilon_t)_{0 \leq t \leq T}$.}
   Initialize imputed values $X^{*(0)}$ and code vectors $W^{(0)}$ \;
   Initialize the counter of iterations: $s=0$ \;
  \While{not converged}{
  Increment $s$: Set $s=s+1$ \;
    Update code vectors by  Kohonen Algorithm  \ref{algo:standardSOM} on the  augmented data with winning neurons obtained by $h^{\mathrm{miss}}$ instead of $h$:\\
  \quad$W^{(s)}\leftarrow \text{Kohonen}(X^{\mathrm{aug}}=(X^{\mathrm{obs}},X^{*(s-1)}))$  \;
  Update  imputed values: for $i,j$ such that $m_{i,j}=0$,
  \begin{equation*}
  x^{*(s)}_{i, j} =
  \frac{\sum_{k=1}^{K} V_{\lambda_T}(k, h^{\mathrm{miss}}(x_i^{\mathrm{obs}},m_i,W^{(s)}))w^{(s)}_{k, j}}{\sum_{k=1}^{K} V_{\lambda_T}(k, h^{\mathrm{miss}}(x_i^{\mathrm{obs}},m_i,W^{(s)}))}.
\end{equation*}
  }
  \KwResult{Code vectors $W^{(s)}$ and   imputed data  $X^{*(s)}$. }
     \caption{missSOM algorithm}\label{algo:missSOM}
\end{algorithm}

\subsection{Minimization algorithm}\label{subsec:missSOM}

For the  minimization of $(W, X^{*})\mapsto F_{\mathrm{miss}}(W, X^{*})$ on $\mathbb{R}^{K\times p}\times  \mathcal{X}^{\mathrm{miss}}$ we propose to alternate the minimization in $W$ and $X^{*}$ while keeping  the other argument
 fixed.

For fixed $X^{*}$, the function $W\mapsto F_{\mathrm{missom}}(W, X^{*})$ is similar to  the objective function $F$ in~\eqref{eq:lossSOM} in the complete-data case applied to the augmented data    $X^{\mathrm{aug}}=(X^{\mathrm{obs}},X^{*})$. The only difference lies in the     definition  of the winning neurons by  $h^{\mathrm{miss}}$   that appear in the neighbourhood function $V_\lambda$. Thus,  a Kohonen algorithm  applied to  $X^{\mathrm{aug}}$  can be used  to find the best code vectors~$W$.

In turn, when $W$ is fixed, the minimization of $ X^{*} \mapsto F_{\mathrm{missom}}(W,  X^{*})$  boils down to minimize $ X^{*} \mapsto F_{\mathrm{miss}}(W,  X^{*})$. 
This problem has a unique  explicit   solution  given  for all $1\le i\le n$ and $j$ such that $m_{i,j}=0$ by
\begin{equation}
x^{*}_{i, j} = \frac{\sum_{k=1}^{K} V_{\lambda}(k, h^{\mathrm{miss}}(x_i^{\mathrm{obs}},m_i,W))w_{k, j}}{\sum_{k=1}^{K} V_{\lambda}(k, h^{\mathrm{miss}}(x_i^{\mathrm{obs}},m_i,W))}.
\label{eq:updateMIssingValues}
\end{equation}
That is, the imputed values are a weighted mean of the  prototype vectors weighted according to the neighborhood function.

To summarize, the algorithm  updates imputed values for the missing data and applies the classical Kohonen algorithm with adjusted winning neuron function $h^{\mathrm{miss}}$ to learn the map. This is repeated until convergence or until a maximum number of iterations chosen by the user is attained. The algorithm is described in Algorithm~\ref{algo:missSOM}.

As initial values for the imputed values $X^*$, one can simply impute the sample mean or median of the variables obtained over the observed entries.

\begin{algorithm}[t]
   \KwData{Incomplete data matrix $X^{\mathrm{obs}}$, missing-data pattern $M$, size and topology of the map,  neighborhood function $V_\lambda$, sequence of radii $(\lambda_t)_{0 \leq t \leq T}$ and learning steps $(\varepsilon_t)_{0 \leq t \leq T}$.}
   Initialize imputed values $X^{*(0)}$ and code vectors $W^{(0)}$ \;
   Initialize the  number of epochs: $t=0$ \;
  \While{not converged}{
  Increment $t$: Set $t=t+1$ \;
  Set $\tilde W^{(0)}=W^{(t-1)}$ \;
  \For{$i=1,\dots,n$}{
    Assignment: Compute winning neuron $\ell=h^{\mathrm{miss}}(x_i^{\mathrm{obs}}, m_i, \tilde W^{(i-1)})$ \;
        Update code vectors:    \\ 
   \For{$k=1,\dots,K$}{$w^{(i)}_k = \tilde w^{(i-1)}_k + 
   \varepsilon_tV_{\lambda_t}(k, \ell) \left((x_i^{\mathrm{obs}}, x_i^{*(t-1)})-\tilde w^{(i-1)}_k\right).$}
   }
   Set $W^{(t)}=\tilde W^{(n)}$\;
    Update  imputed values: for $i,j$ such that $m_{i,j}=0$,
\begin{equation*}  x^{*(t)}_{i, j} =
\frac{\sum_{k=1}^{K} V_{\lambda_t}(k, h^{\mathrm{miss}}(x_i^{\mathrm{obs}},m_i,W^{(t)}))w^{(t)}_{k, j}}{\sum_{k=1}^{K} V_{\lambda_t}(k, h^{\mathrm{miss}}(x_i^{\mathrm{obs}},m_i,W^{(t)}))}.
\end{equation*}
  }
  \KwResult{Code vectors $W^{(t)}$ and   imputed data  $X^{*(t)}$. }
     \caption{Accelerated missSOM algorithm}\label{algo:accelmissSOM}
\end{algorithm}

 \subsection{Accelerated version}\label{subsec:missSOMaccel}

Algorithm~\ref{algo:missSOM} happens to  be expensive in terms of computing time, in particular when the number of iterations is large, since  the entire standard Kohonen algorithm  is carried out during each iteration. 
A speed up is obtained by interwining updates of the missing data and the iterations of the Kohonen  Algorithm \ref{algo:standardSOM}.  
More precisely, we propose to update the missing data at every epoch, that is, after every pass through the data. 
This procedure gives rise to   Algorithm~\ref{algo:accelmissSOM}. 
 As such, the Kohonen algorithm is carried out only once, while in the initial Algorithm~\ref{algo:missSOM}, the entire Kohonen algorithm is applied repeatedly. Thus the computing time of the accelerated version of missSOM is comparable to the computing time of the standard Kohonen Algorithm \ref{algo:standardSOM}, since the update of the imputed values is fast.

 While the first version of the missSOM algorithm  has some  theoretical justification, the accelerated version lacks this foundation. A numerical study given in Appendix ~\ref{subsec:simul-theo-accel} shows that the Algorithms 
\ref{algo:missSOM}  and \ref{algo:accelmissSOM} provide very similar maps and hence justifies the utilization of the accelerated version, which achieves a significant gain in computing time.

 Note that while the selection of the observations $x_i$ in Algorithm~\ref{algo:accelmissSOM} is deterministic, it is  possible to use a  random selection scheme.

\section{Numerical experiments} \label{sec:numericalStudy}

In this section, the performance of  the proposed method missSOM   is evaluated and compared to alternative  methods. 
A simulation study is conducted to  assess the quality of the   representation of the data via the map  and the  accuracy of imputed values under various conditions.
 
\subsection{Performance criteria}

Let  $W^*$ be  the   code vectors of the final self-organizing map, $X^*$ the imputed values and $\hat x_i=(x_i^\mathrm{obs},x_i^*)$ the completed observation vectors.

The quality of the map as a representation of the data 
  can be evaluated by two criteria. 
First, the {\it quantization error} defined as the average of the squared distances between the observations and their nearest prototype vector given by 

$$E = \frac{1}{n_{\mathrm{obs}}}\sum_{i=1}^{n}\|x_i^\mathrm{obs}-w_{h^{\mathrm{miss}}(x_i^{\mathrm{obs}},m_i,W^*)}\|^2_2,$$
where $n_{\mathrm{obs}}=\sum_{i, j}m_{i, j}$ is the number of observed entries in the data, informs on whether the prototype vectors are good representations of the data. 
Second, the {\it topographic error}  evaluates the  preservation of the topology of the data in the map by the proportion of observations for which the winning neuron and   the  second closest neuron are not neighbors, i.e.   not connected on the grid. It is defined as
    \begin{equation*}
T = \frac{1}{n}\sum_{i=1}^{n} e(\hat x_i),
\end{equation*}
where
\begin{equation*}
e(\hat x_i) = \left\{
\begin{array}{ll}
0 & \mbox{if the neurons ${h^{\mathrm{miss}}(x_i^{\mathrm{obs}},m_i,W^*)}$ and $v^{\mathrm{miss}}(x_i^{\mathrm{obs}},m_i,W^*)$ are adjacent}
  \\
1 & \mbox{otherwise }
\end{array}
\right.
\end{equation*}
with $v^{\mathrm{miss}}(x_i^{\mathrm{obs}},m_i,W^*) = \arg\min_{j \neq h^{\mathrm{miss}}(x_i^{\mathrm{obs}},m_i,W^*) }\|x_i^{\mathrm{obs}} -w^*_{k}\odot m_i\|^2_2$ the second closest neuron to measurement  $x_i^{\mathrm{obs}}$.
The topographic error is expected to be small when the map is well organized and ordered. 
As both errors are computed only on the observed entries,  the focus is put on   the quality of the map with respect to the observed part of the data.  The accuracy of the imputed values is assessed separately by the 
 {\it imputation error}, which  
is defined as the root mean square error and quantifies the quality of the imputed values compared to the true missing values.

\subsection{Datasets}

For the numerical experiments,   two settings are considered. First, the   dataset {\it wines}   from the UCI machine learning repository \citep{Dua:2019} is used, which contains the results of a chemical analysis of 178 wines on 13 quantitative variables. 
We generate  100 perturbed datasets by adding   gaussian noise with mean 0 and standard deviation equal to one tenth of the mean value in each variable. 

In the second setting,  100 datasets are simulated from a multivariate gaussian mixture 
with dimension $p=5$, four equal-sized groups and  $n=2000$ observations.  The correlation among all pairs of  components is equal to 0.5 and
for every dataset, the 4 gaussian means are drawn independently from a centered normal distribution with  standard deviation equal to 5.

In both settings,  missing values are generated using the \texttt{ampute} function from the R package \texttt{mice}~\citep{mice}. Different proportions of missing values, namely  5\%, 20\% and 40\%, and    different mechanisms of missingness are considered. 
The literature on missing data traditionally distinguishes  
three mechanisms that lead to missingness~\cite{rubin} and it is well known that the performance of imputation methods may be sensitive to the  mechansim at work. 
The mechanism is said to be {\it missing completely at random} (MCAR), when the causes of missing values are independent from the data and the probability of being absent is  the same for all items. 
That is, 
a subset of observations is chosen at random  using independent  Bernoulli variables with fixed success probability for all entries.
 In contrast, when the probability that a value is missing depends   on the values of the observed variables, the mechanism is called {\it missing at random} (MAR). In our simulations, to obtain MAR, for each variable,  
 missing values are obtained using a logistic regression model depending only on the other variables.
Finally,  when the probability of being absent also depends on the unobserved 
value, 
the mechanism is called {\it missing not at random} (MNAR) and  we can consider missing data simulated by using a logistic regression model depending on all variables.

\subsection{Alternative methods}

Our study is twofold. On the one hand, we compare {\it missSOM} to the state of the art on self-organizing maps. The simplest method (referred to  as {\it deletion} in the figures) consists in deleting the observations containing missing values and applying the standard Kohonen algorithm to the remaining data. Incomplete observations are classified once the map is built by assigning them to their closest prototypes 
and missing entries are imputed by the corresponding values of the winning prototypes.  {\it Cottrell}'s approach  is a variant of  the classical self-organizing maps
 appropriate to  deal with incomplete observations~\cite{Cottrell2007MissingV}.
 {\it Cottrell}
  adapts the Kohonen algorithm by simply restricting all vector calculations  to the observed entries.  
 The main difference with \textit{missSOM} is that data imputation is performed only after learning the  map by imputation with the values of the closest prototypes.
 
%
 
On the other hand, in our experiments, we   compare {\it missSOM} to the state of the art on missing data imputation. 
The literature provides numerous general imputation methods.
A basic approach (here referred to as  {\it mean}) is the imputation by the mean value of the observed variables. 
A non-parametric approach called {\it missForest}    predicts  missing values using a random forest trained on the observed part of the dataset.
 Moreover, the method   referred to as  {\it knn} is a $k$-nearest neighbors approach, which is  implemented in the \texttt{VIM} package~\citep{vim}. Missing values are    imputed  iteratively    by a weighted average of the $k$ closest observations. 
Finally, assuming a gaussian mixture model for the data, the \texttt{Amelia} package~\citep{amelia} performs imputation using the expectation-maximization  algorithm and  a bootstrap approach to iteratively estimate missing values. In our numerical study,   the imputation error of missSOM is compared to the one obtained by all these methods. In addition, we also  compare the map obtained by missSOM with the maps obtained by the classical Kohonen algorithm   applied to the complete dataset, where missing values are imputed by the aforementioned imputation methods.


As a benchmark we consider a self-organizing map trained on  the complete data set (corresponding to a missing rate of 0\%) and compute the error rates only on the observed values that are provided to the other methods. This method is referred to as \textit{complete-data SOM}. 

\subsection{Parameters of the methods}

The missSOM algorithm is applied with the default  parameters in the package \texttt{missSOM}. That is, the neighborhood function is a   gaussian and  the sequence $(\lambda_t)_{t \leq T}$ decreases from $\lambda_0=4.58$ to $\lambda_T=0.5$ for the wine dataset and from $\lambda_0=8.89$ to $\lambda_T=0.5$ for the gaussian mixture data. The maximum number of epochs is $T=100$.
Concerning  the map, it  has a hexagonal topology and the    grid is composed of $K = 9 \times 7$ neurons for the wines data and $K=16 \times 14$ neurons for the gaussian mixture data. 
The choice of the size of the map depends on the sample size and on the objectif of the analysis.
 In general, missSOM has the same parameters as classical SOM and they can be chosen in the same way as for SOM.


 For the alternative methods, the maximum number of iterations is set to $100$ and for the other parameters the default values are used.

\subsection{Comparison to the state of the art on self-organizing maps}

\begin{figure}[!tbp]
  \makebox[\textwidth][c]{%
  \subfloat[Wines data]{\scalebox{0.3}{\includegraphics{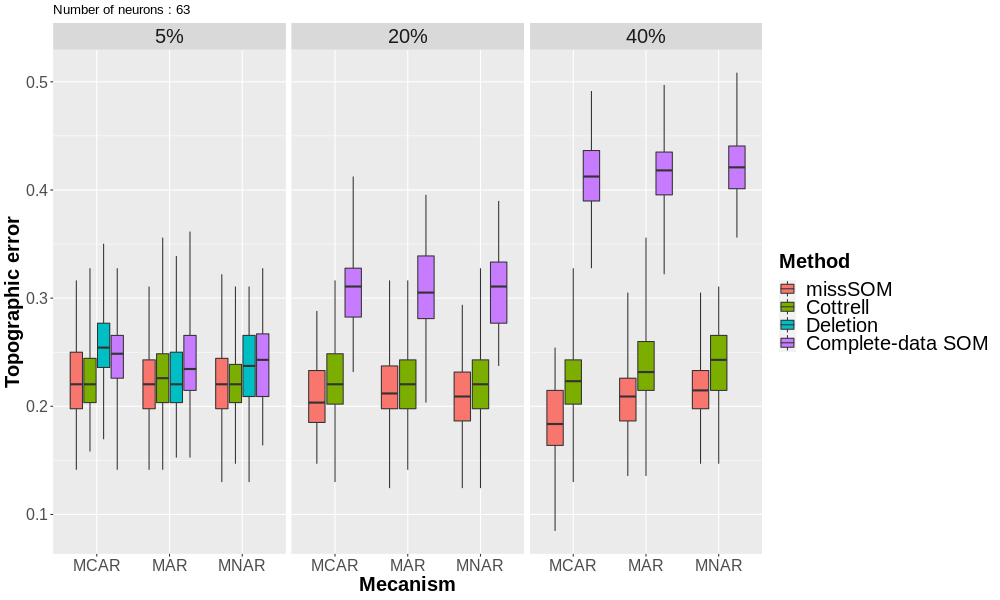}}}
  \\
  \subfloat[Gaussian mixture data]{\scalebox{0.3}{\includegraphics{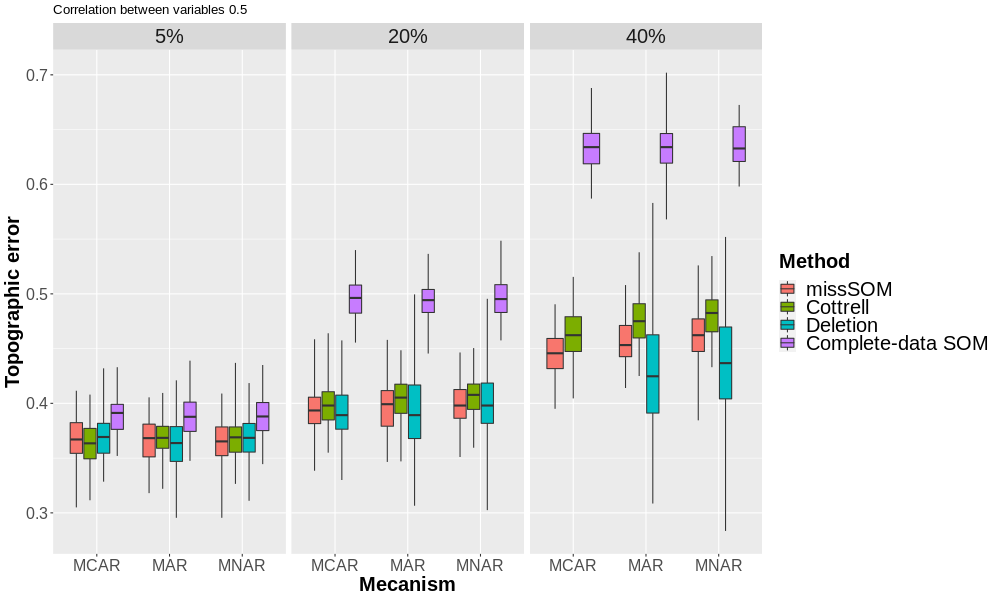}}}
  }
   \caption{Topographic error of SOM-type methods for various amounts and   mechanisms of missingness on the wines (a) and the gaussian mixture data (b). }
  \label{fig:som:topo}
\end{figure}

\begin{figure}[!tbp]
  \makebox[\textwidth][c]{%
  \subfloat[Wines data]{\scalebox{0.3}{\includegraphics{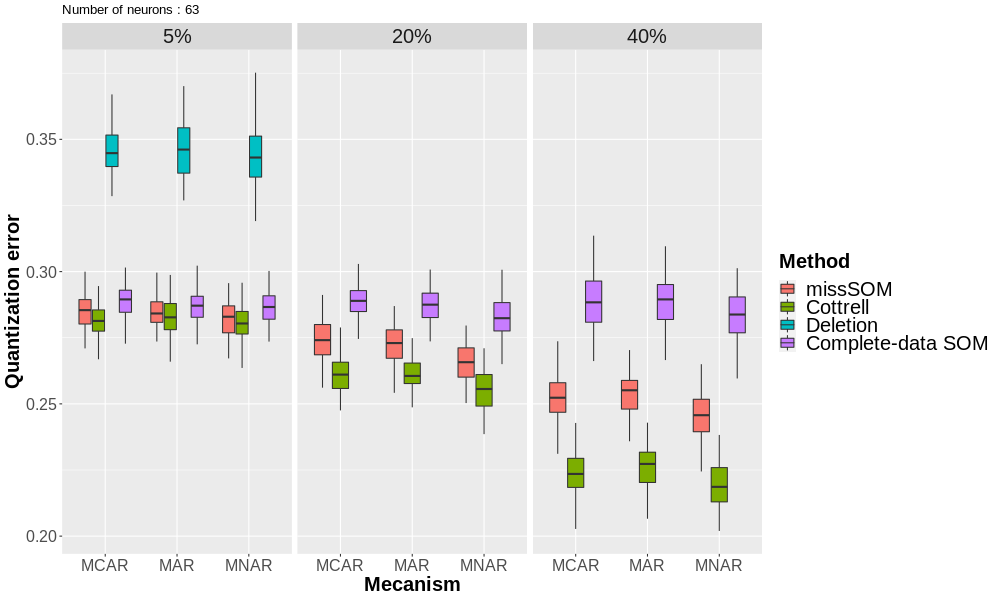}}\label{fig:f4}}
  \quad
  \subfloat[Gaussian mixture data]{\scalebox{0.3}{\includegraphics{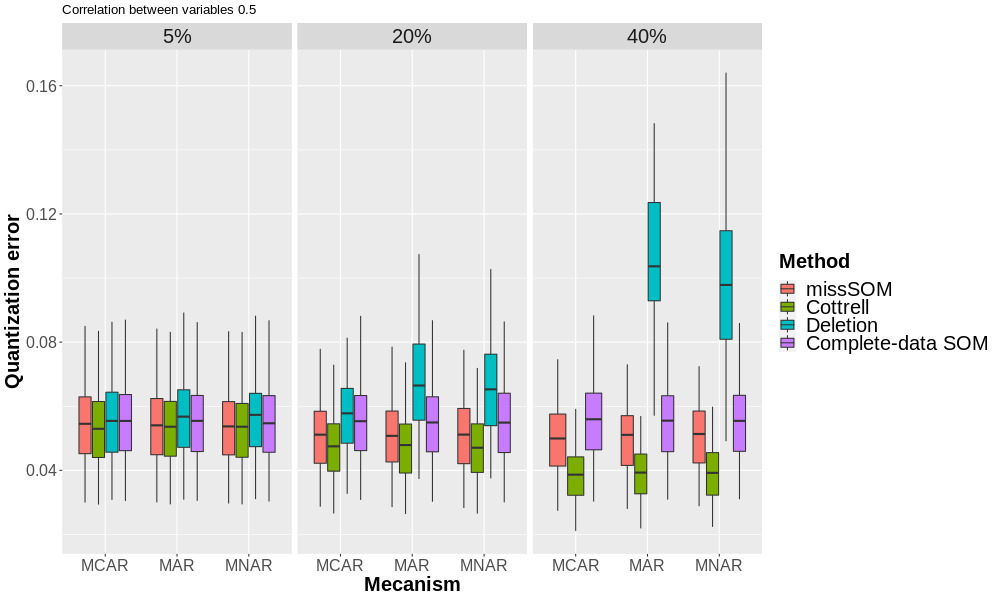}}\label{fig:f5}}
  }
  \caption{Quantization error  of SOM-type methods for various amounts and   mechanisms of missingness on the wines (a) and the gaussian mixture data (b).}
  \label{fig:som:quantif}
\end{figure}

\begin{figure}[!tbp]
  \makebox[\textwidth][c]{%
  \subfloat[Wines data]{\scalebox{0.3}{\includegraphics{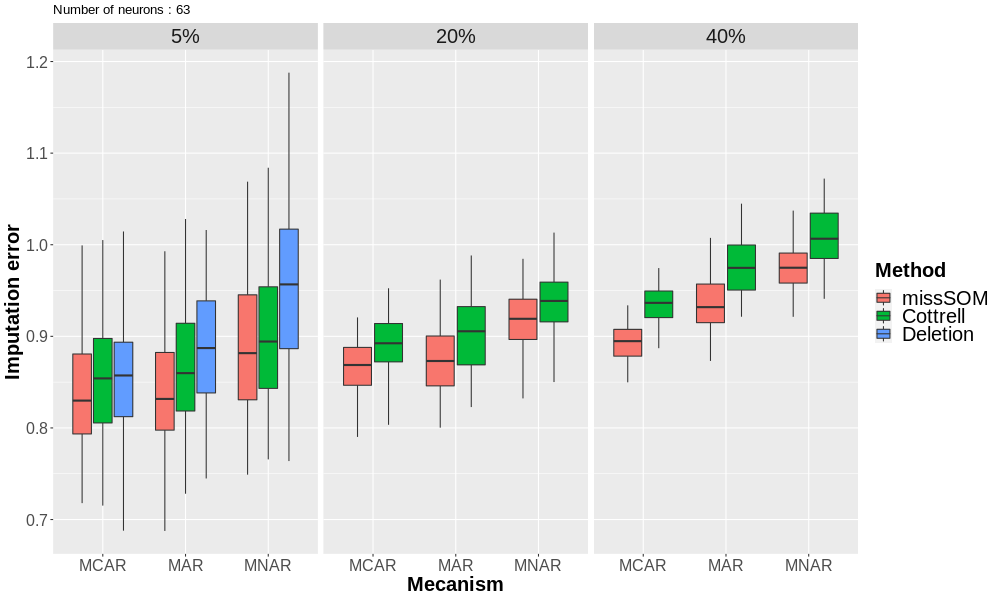}}\label{fig:f4}}
  \quad
  \subfloat[Gaussian mixture data]{\scalebox{0.3}{\includegraphics{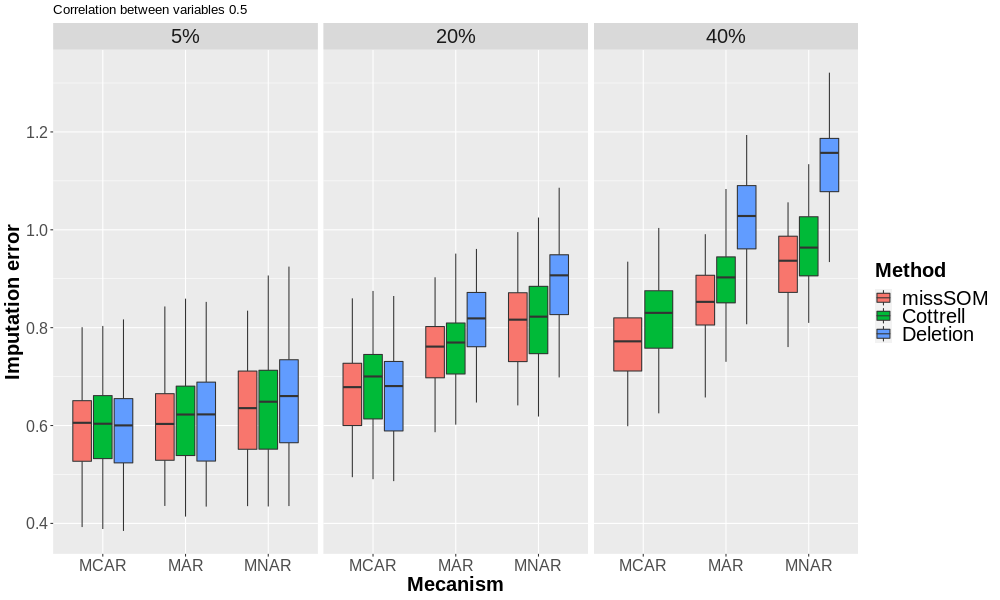}}\label{fig:f5}}
  }
  \caption{Imputation error   of SOM-type methods for various amounts and  mechanisms of missingness on the wines (a) and the gaussian mixture data (b).
  }
  \label{fig:som:impute}
\end{figure}

Figure \ref{fig:som:topo} and \ref{fig:som:quantif} present the boxplots of  the topographic   and the quantization error for all SOM-type  methods in both settings.  They serve to evaluate the quality of the representation of the data using a self-organizing map computed with the different approaches. The corresponding imputation errors in Figure \ref{fig:som:impute} allow to judge their performances as   imputation methods.

First, we observe that when the  percentage of missing data is low, most methods have very similar performance. With increasing percentage of missingness, the problem becomes harder and differences among the methods appear. But as errors are evaluated only on the observed part of the data, errors are not necessarily increasing. 
Namely  quantization errors for the wines data decrease with increasing missingness. This may be a consequence of the higher variance and less structure of  the wines data  compared to the Gaussian mixture data. So when deleting entries from the wines data, the data variance decreases and the resulting map is a better representation of the observed part of the data.
Concerning the mechanisms of missingness, it appears that the quality of the map 
  does not depend on it for any SOM-type  method. However, imputation is impacted by the mechanism. Imputation  is  the easiest under MCAR and the hardest under MNAR.

Next, we see that the {\it deletion} method is very unstable. In some scenarios its error rates are among the worst, and, more importantly, when too many  values are missing, the method breaks down and does not produce any result. Indeed, on small datasets as wines with 20\% of missingness, the number of complete observations  is smaller than the size of the map and thus classical SOM is just not applicable. On the large gaussian mixture dataset,  the {\it deletion} method produces the best topographic error, but the associated quantization error is disastrous, disqualifying the approach.  
 
 The quantization error of the  {\it complete-data SOM} method is constant when varying the amount of missingness, because the underlying map remains the same. For the topographic error, an increase is observed which is due to the definition of the error.  The error  determines   the closest and second closest prototypes only with respect to the observed entries, while the map was optimized by taking into account the complete data. This explains why  {\it missSOM} and {\it Cottrell}  have lower topographic errors, as their maps are learned with a notion of closeness restricted on the observed entries.

Finally, we observe that  {\it Cottrell}'s method always achieves the best quantization errors, directly followed by {\it missSOM}.
Concerning  the   topographic error, {\it missSOM} is consistently doing better than   {\it Cottrell}. 
Thus, in terms of quality of the map and representation of  the data, none of the methods outperforms all others, and   {\it Cottrell} and {\it missSOM} have both a good global performance.
 Now, considering the  imputation error, there is a clear winner. When the proportion of missingness is important, {\it missSOM} outperforms all other methods in every scenario.
  This  confirms us in the use of {\it missSOM} with respect to {\it Cottrell}, especially when accurate imputation is desired. 
 
\subsection{Comparison to the state of the art on missing-data imputation}

\begin{figure}[!tbp]
  \makebox[\textwidth][c]{%
  \subfloat[Wines data]{\scalebox{0.3}{\includegraphics{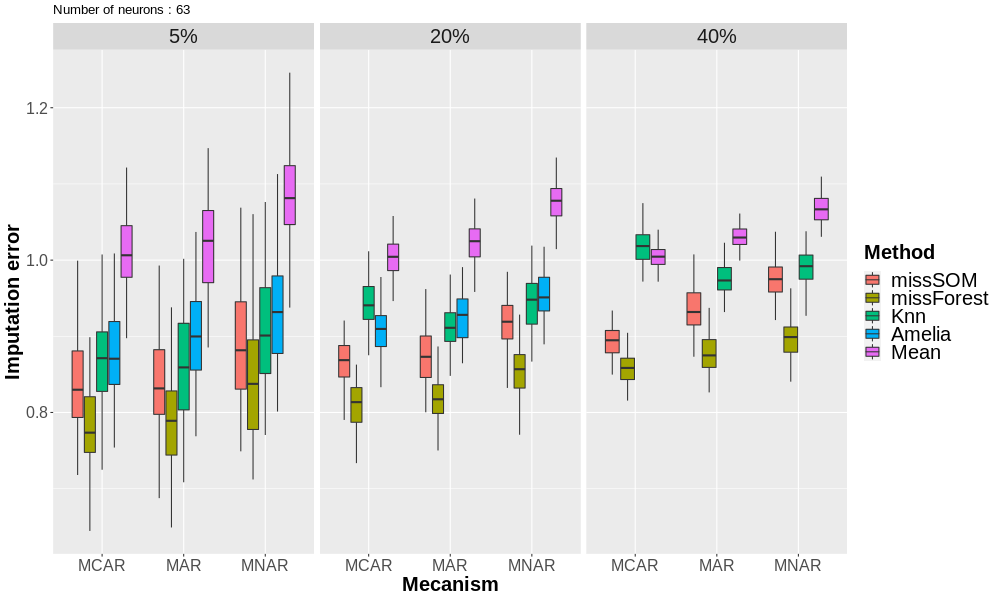}}\label{fig:f4}}
  \quad
  \subfloat[Gaussian mixture data]{\scalebox{0.3}{\includegraphics{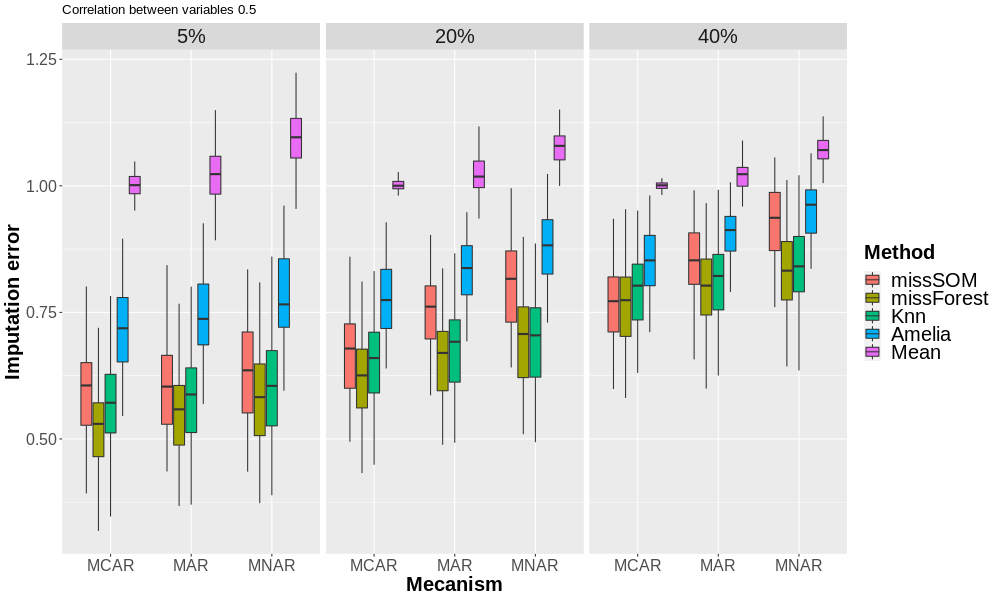}}\label{fig:f5}}
  }
        \caption{Imputation error   of  {\it missSOM} and classical imputation methods for various amounts and  mechanisms of missingness on the wines (a) and the gaussian mixture data (b).
  }
  \label{fig:imput:impute}
\end{figure}

\begin{figure}[!tbp]
  \makebox[\textwidth][c]{%
  \subfloat[Wines data]{\scalebox{0.3}{\includegraphics{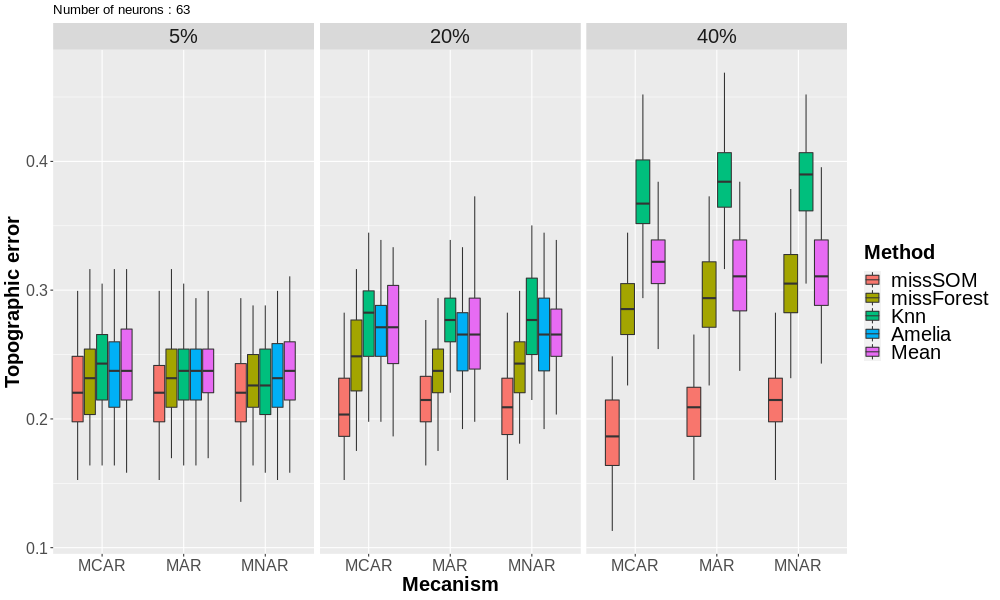}}\label{fig:f4}}
  \quad
  \subfloat[Gaussian mixture data]{\scalebox{0.3}{\includegraphics{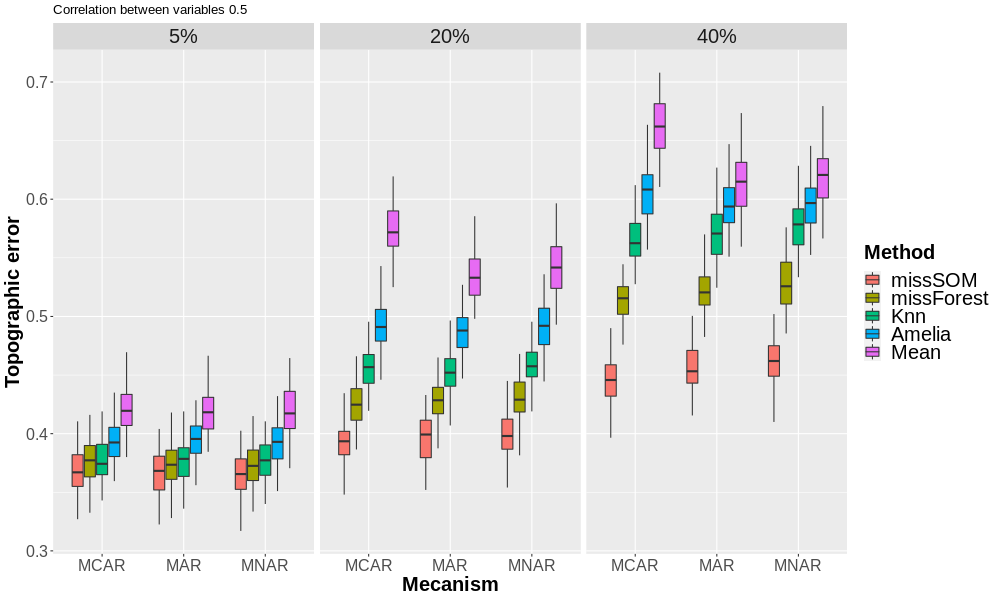}}\label{fig:f5}}
  }
    \caption{Topographic error   of  {\it missSOM} and classical imputation methods for various amounts and  mechanisms of missingness on the wines (a) and the gaussian mixture data (b).
  }
  \label{fig:imput:topo}
\end{figure}

\begin{figure}[!tbp]
  \makebox[\textwidth][c]{%
  \subfloat[Wines data]{\scalebox{0.3}{\includegraphics{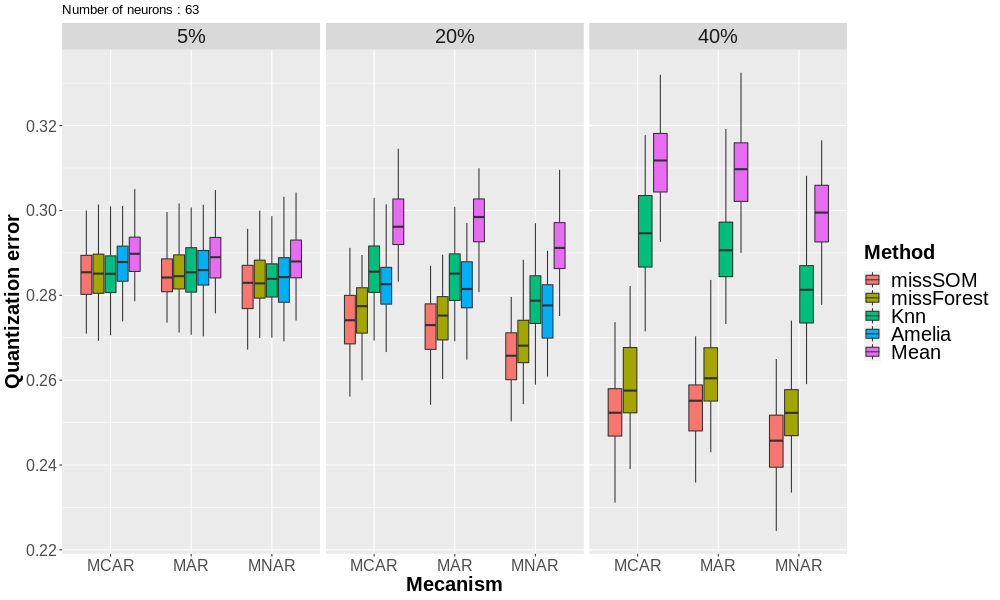}}\label{fig:f4}}
  \quad
  \subfloat[Gaussian mixture data]{\scalebox{0.3}{\includegraphics{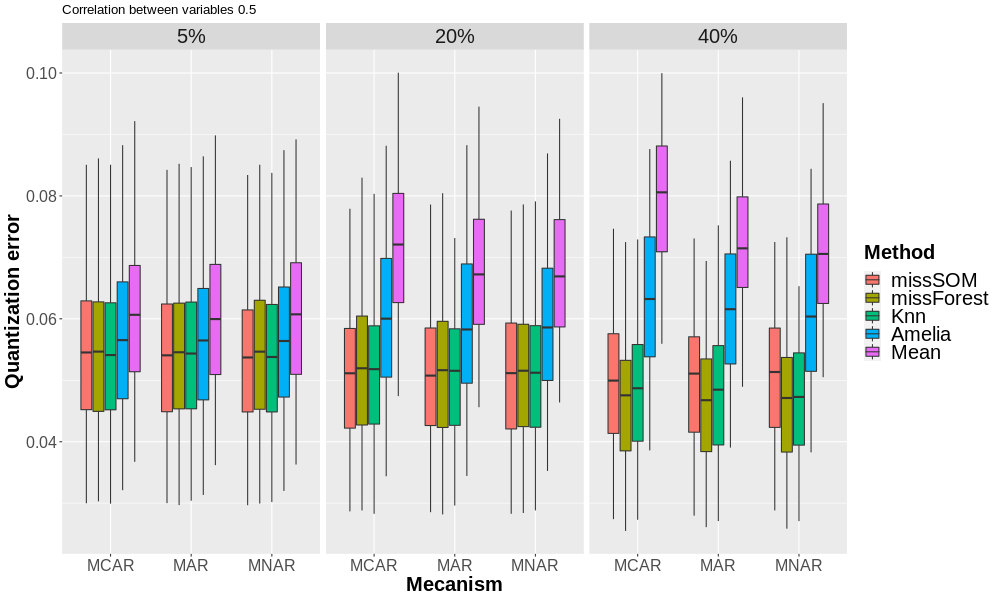}}\label{fig:f5}}
  }
      \caption{Quantization error   of  {\it missSOM} and classical imputation methods for various amounts and  mechanisms of missingness on the wines (a) and the gaussian mixture data (b).
  }

  \label{fig:imput:quantif}
\end{figure}

In the second part of the simulation study,  {\it missSOM}  is    compared to general imputation methods.   
Figure \ref{fig:imput:impute} shows   that as an imputation method {\it missForest} is unbeaten regardless of the missingness mechanism and the percentage of missing data. On gaussian mixture data, {\it knn} also provides accurate imputations.
The imputation accuracy achieved by {\it missSOM} is high and   {\it missSOM}  is mostly among the best imputation methods.

Finally, Figure \ref{fig:imput:topo} and Figure \ref{fig:imput:quantif} provide insights on the quality of the data representation, when   classical SOM is applied to the data   with missing values imputed by the different imputation methods. In this respect, {\it missSOM} clearly outperforms all the other methods. In particular,  {\it missForest} provides self-organizing maps of poorer  quality,  especially in terms of preservation of the topology of the input data. 
 This confirms the initial motivation of {\it missSOM}: in the presence of missing values and when a self-organizing map is required, it is better to  learn the map simultaneously with the imputed values than to treat the two tasks separately. 
 
A further advantage of  {\it missSOM} is its computing time, which is about 10 times shorter than the one of {\it missForest}. 

   To summarize, we have seen that   
    {\it missSOM} is the method of choice when both tasks data imputation and data representation are desired. However, when the focus is only on data imputation, then better alternatives as {\it missForest} exist.


\section{Conclusion}

In this paper, we have proposed an extension of the self-organizing map for partially observed data, referred to as  missSOM. The proposed method addresses simultaneously the two problems of computing a self-organizing map and imputing missing values. 
A numerical study assesses the good performance of 
missSOM regarding various criteria and in comparison to the state of the art.
While this paper focuses on the standard Kohonen algorithm, in future work we may address the  
 transfer of our  approach to other existing  variants of self-organizing maps on more complex data types such as mixed data to enable them to deal with missing data.


\section{Appendix}

\subsection{Validation of accelerated missSOM algorithm}\label{subsec:simul-theo-accel}

Table \ref{tab:version_algo_topo_error} compares the results of the basic missSOM Algorithm \ref{algo:missSOM} and its accelerated version   Algorithm \ref{algo:accelmissSOM}  on the simulated gaussian mixture data in various conditions. On the one hand, for all settings the errors are totally equivalent. This indicates that the accelerated version provides the same self-organizing maps and very similar imputations as the basic missSOM algorithm. On the other hand, we see that in terms of computing time we gain two orders of magnitude. Hence,  the use of the accelerated algorithm instead of the basic version is completely justified.
 
\begin{table*}[t]
    \centering
    \begin{adjustbox}{max width=15cm}
    \begin{tabular}{lccc  |ccc  | ccc}
    
 & \multicolumn{3}{c|}{5\%} &  \multicolumn{3}{c|}{20\%} &   \multicolumn{3}{c}{40\%}\\
    & MCAR  & MAR  & MNAR  & MCAR & MAR  & MNAR & MCAR & MAR  & MNAR \\
    \hline
{\bf Topographic error}&&&&&&&&&\\
basic missSOM        & 0.337 (0.047) & 0.345 (0.035) & 0.336 (0.037)   &
              0.313 (0.032) & 0.355 (0.063) & 0.339 (0.046)   &
              0.294 (0.058) & 0.303 (0.057) & 0.306 (0.083)  \\
accelerated missSOM  & 0.352 (0.039) & 0.345 (0.049) & 0.338 (0.030)   &
              0.301 (0.035) & 0.299 (0.039) & 0.322 (0.042)   &
              0.282 (0.035) & 0.280 (0.059) & 0.285 (0.040)   \\
              \hline
{\bf Quantization error}&&&&&&&&&\\
basic missSOM        & 0.405 (0.115) & 0.394 (0.042) & 0.407 (0.145)   &  
              0.327 (0.115) & 0.326 (0.116) & 0.330 (0.112)   &
              0.225 (0.074) & 0.233 (0.080) & 0.233 (0.077)  \\
                            accelerated missSOM     & 0.413 (0.149) & 0.408 (0.139) & 0.407 (0.138)   &
              0.351 (0.117) & 0.340 (0.116) & 0.341 (0.115)   &
              0.261 (0.081) & 0.245 (0.071) & 0.260 (0.074)  \\
               
\hline
{\bf Imputation error}&&&&&&&&&\\
basic missSOM  & 0.601 (0.140) & 0.599 (0.154) & 0.613 (0.171)   &
              0.659 (0.125) & 0.686 (0.141) & 0.759 (0.108)   &
              0.758 (0.093) & 0.795 (0.095) & 0.871 (0.091)   \\
accelerated missSOM       & 0.615 (0.097) & 0.624 (0.096) & 0.654 (0.103)   &
              0.665 (0.092) & 0.694 (0.087) & 0.750 (0.075)   &
              0.742 (0.069) & 0.793 (0.069) & 0.848 (0.072)  \\
              \hline  
{\bf ARI}&&&&&&&&&\\
basic missSOM         & 0.922 (0.092) & 0.929 (0.089) & 0.937 (0.071)   &
              0.849 (0.088) & 0.846 (0.108) & 0.823 (0.089)   &
              0.686 (0.108) & 0.640 (0.101) & 0.623 (0.086)  \\
accelerated missSOM        & 0.923 (0.090) & 0.923 (0.086) & 0.922 (0.082)   &
              0.841 (0.092) & 0.845 (0.081) & 0.805 (0.090)   &
              0.647 (0.094) & 0.629 (0.109) & 0.591 (0.089)   \\
              \hline
{\bf Computing time}&&&&&&&&&\\
basic missSOM        & 33.169 (0.546) & 32.940 (0.737) & 31.860 (0.396)   &
              33.076 (0.517) & 32.988 (0.776) & 32.948 (0.530) &
              33.031 (0.583) & 31.920 (0.061) & 32.846 (0.0644)   \\
accelerated missSOM        & 0.394 (0.010) & 0.387 (0.006) & 0.376 (0.009)   &
              0.557 (0.008) & 0.561 (0.017) & 0.556 (0.006)   & 
              0.790 (0.016) & 0.753 (0.011) & 0.784 (0.022)  \\
        \end{tabular}
                                              
    \end{adjustbox}
    \caption{Comparison of the basic missSOM Algorithm \ref{algo:missSOM} and its accelerated version   Algorithm \ref{algo:accelmissSOM} in terms of different errors and computing time (in seconds) on the gaussian mixture data.}
    \label{tab:version_algo_topo_error}
                                          
\end{table*}


\bibliography{mybibfile_New}

\end{document}